\documentclass[prl,twocolumn]{revtex4}
 \usepackage[english]{babel}
 \usepackage{epsfig,amssymb,amsmath}
 \usepackage{color}

\def\work{Letter}

\begin{document}

\title{Nonequilibrium resonant spectroscopy of molecular vibrons}

\author{Dmitry~A.~Ryndyk and Gianaurelio~Cuniberti}

\affiliation{Institut f\"ur Theoretische Physik, Universit\"at Regensburg, D-93040 Germany}

\begin{abstract}
Quantum transport through single molecules is essentially affected by molecular
vibrations. We investigate the behavior of the prototype single-level model with
intermediate electron-vibron coupling and arbitrary coupling to the leads. We have
developed a theory which allows to explore this regime via the nonequilibrium Green
function formalism. We show that the nonequilibrium resonant spectroscopy is able to
determine the energies of molecular orbitals and the spectrum of molecular vibrations.
Our results are relevant to scanning tunneling spectroscopy experiments, and demonstrate
the importance of the systematic and self-consistent investigation of the effects of the
vibronic dynamics onto the transport through single molecules.
\end{abstract}

\date{\today}
\maketitle

Molecular electronics \cite{Reed00sciam,Joachim00nature,Cuniberti05book} is the most
promising development of nano-electronics, raising new theoretical challenges and calling
for new methods. The understanding of quantum electron transport at the molecular scale
is a key step to future devices. Recently, the interaction of electrons with molecular
vibrations attracted an enormous attention unveiled by electron transport experiments
through single molecules
\cite{Reed97science,Park00nature,Park02nature,Liang02nature,Smith02nature,Yu}. In
scanning tunneling spectroscopy (STS) experiments clear signatures of the electron-vibron
interaction have been observed \cite{Qiu04prl,Wu04prl,Repp05prl,Repp05prl2}. In these
experiments both the electron-vibron coupling and the electron hybridization are well
beyond the perturbation limit, so that a theory beyond the linear response or master
equation approaches is necessary. In this {\work}, we present such a theory and discuss
the vibronic features in the current-voltage curves of molecular junctions as a
spectroscopic tool.

\begin{figure}[b]
\begin{center}
\epsfxsize=0.9\hsize \vskip 0.5cm
\epsfbox{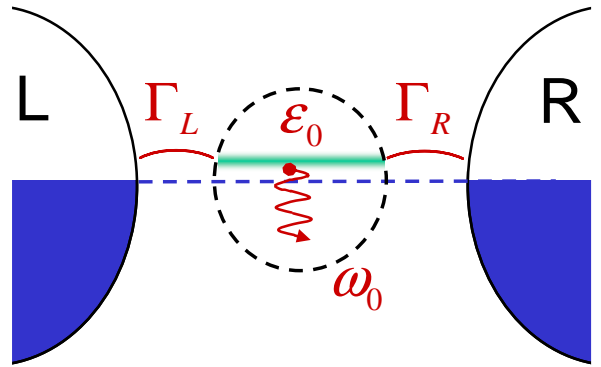}
\caption{(Color online) Schematic picture of the considered fluctuating single-level model.}
\label{fig1}
\end{center}
\end{figure}

Resonant electron transport through a molecular orbital with electron-vibron interaction
at finite voltages is essentially different from the well known inelastic electron
tunneling spectroscopy (IETS). Indeed, in the IETS situation one has typically the
multi-channel tunneling between two bulk metals with continuous density of states, and an
{\em increase} of the conductance is observed when the bias voltage exceeds the threshold
determined by the frequency of vibrational mode. As we shall see below, in resonant
tunneling through molecular levels coupled to vibrons, the situation is very different:
in addition to the usual IETS signal, the inelastic signal can be observed as an extra
{\em peak} in the differential conductance, or just as a {\em decrease} of the
conductance. The resulting signal is determined by the energy of the electronic state,
the vibronic frequency, the electron-vibron coupling, and the molecule-to-lead coupling.
The position of the molecular energy levels could be changed by the external gate
voltage, but in the STS experiments the application of the gate voltage is not possible,
instead the tip-to-molecule distance can be changed, also changing the molecule-to-tip
coupling \cite{Wu04prl}. Thus it is very important to investigate quantum transport at
arbitrary coupling to the leads. The simplest model, which describes properly these
peculiarities, is a single electron with energy $\epsilon_0$ interacting with a single
vibron with frequency $\omega_0$ (Fig.\,\ref{fig1}), and coupled to two noninteracting
equilibrium electrodes.

We analyze here the electron-vibron model developing the preliminary results obtained in
our previous publications \cite{Ryndyk05prb,Ryndyk06prb}. We consider a molecule coupled
to free conduction electrons in the leads by a usual tunneling Hamiltonian. Furthermore,
the electrons are coupled to vibrational modes. The full Hamiltonian is the sum of the
molecular Hamiltonian $\hat H_M$, the Hamiltonians of the leads $\hat H_{R(L)}$, the
tunneling Hamiltonian $\hat H_T$ describing the molecule-to-lead coupling, the vibron
Hamiltonian $\hat H_V$ including electron-vibron interaction and coupling of vibrations
to the environment (describing dissipation of vibrons)
\begin{equation}\label{H}
 \hat H=\hat H_M+\hat H_V+\hat H_L+\hat H_R+\hat H_T.
\end{equation}

A molecule is described by a set of localized states $|\alpha\rangle$ with energies
$\epsilon_\alpha$ and inter-orbital overlap integrals $t_{\alpha\beta}$ by the following
model Hamiltonian:
\begin{equation}\label{H_M}
 \hat H^{(0)}_M=\sum_\alpha\left(\epsilon_\alpha+e\varphi_\alpha(t)\right)
 d^{\dag}_\alpha d_\alpha +\sum_{\alpha\neq\beta}t_{\alpha\beta}
 d^{\dag}_\alpha d_\beta,
\end{equation}
where $d^{\dag}_\alpha$,$d_\alpha$ are creation and annihilation operators in the states
$|\alpha\rangle$, and $\varphi_\alpha(t)$ is the (self-consistent) electrical potential.
The index $\alpha$ is used to mark single-electron states (atomic orbitals) including the
spin degree of freedom. The parameters of a tight-binding model could be determined by
{\em ab initio} methods \cite{Cuniberti05book}. This is a compromise, which allows us to
consider complex molecules with a relatively simple model.

The isolated single-level electron-vibron model is described by the Hamiltonian
($\hbar=1$)
\begin{equation}
  \hat H_{M+V}=(\epsilon_0+e\varphi_0)d^{\dag}d+\omega_0a^{\dag}a+\lambda\left(a^{\dag}+a\right)d^{\dag}d,
\end{equation}
where the first and the second terms describe the free electron state and the free
vibron, and the third term is electron-vibron minimal coupling interaction.

The electrical potential of the molecule $\varphi_0$ plays an important role in transport
at finite voltages. It describes the shift of the molecular level by the bias voltage,
which is divided between the left lead (tip), the right lead (substrate), and the
molecule as $\varphi_0=\varphi_R+\eta(\varphi_L-\varphi_R)$ \cite{Datta97prl}.

The Hamiltonians of the right (R) and left (L) leads read
\begin{equation}
 \hat H_{i=L(R)}=\sum_{k\sigma}(\epsilon_{ik\sigma}+e\varphi_i)
 c^{\dag}_{ik\sigma}c_{ik\sigma},
\end{equation}
$\varphi_i(t)$ are the electrical potentials of the leads. Finally, the tunneling
Hamiltonian
\begin{equation}\label{H_T}
 \hat H_T=\sum_{i=L,R}\sum_{k\sigma,\alpha}\left(V_{ik\sigma,\alpha}
 c^{\dag}_{ik\sigma}d_\alpha+{\rm h.c.}\right)
\end{equation}
describes the hopping between the leads and the molecule. A direct hopping between
two leads is neglected.

\begin{figure}
\begin{center}
\epsfxsize=1.0\hsize
\epsfbox{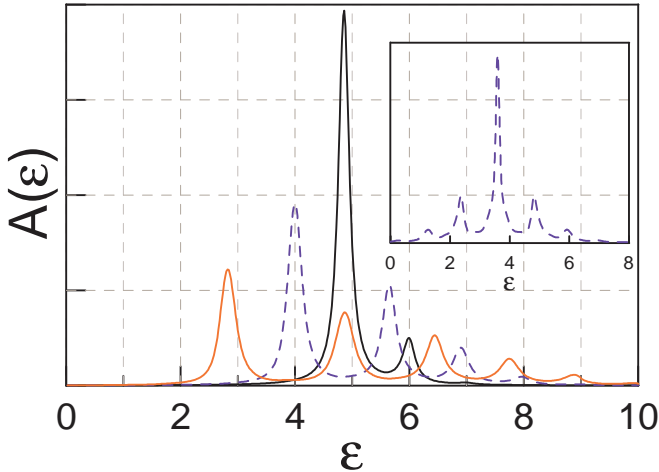}
\caption{(Color online) Spectral function at different electron-vibron couplings: $\lambda/\omega_0=0.4$
(black), $\lambda/\omega_0=1.2$ (blue, dashed), and $\lambda/\omega_0=2$ (red); at $\epsilon_0/\omega_0=5$,
$\Gamma_L/\omega_0=\Gamma_R/\omega_0=0.1$.  In the insert the spectral function at
$\lambda/\omega_0=1.2$ is shown at finite voltage, when the level is partially filled.
Energies are in units of $\hbar\omega_0$.}
\label{fig2}
\end{center}
\end{figure}

Though the model described above has a long history, the many questions are not answered
up to now. While the isolated electron-vibron model can be solved exactly by the
so-called polaron or Lang-Firsov transformation, the coupling to the leads produces a
true many-body problem. The inelastic resonant tunneling of electrons coupled to phonons
was first considered in
Refs.~\cite{Glazman88jetp,Wingreen88prl,Wingreen89prb,Jonson89prb}. There the exact
solution in the single-particle approximation was derived, ignoring completely the Fermi
sea in the leads. At strong electron-vibron couplings and weak couplings to the leads the
satellites of the main resonant peak are formed in the spectral function
(Fig.\,\ref{fig2}). The question which remains is whether these side-bands can be
observed in the differential conductance. New theoretical treatments were presented in
Refs.\,\cite{Lundin02prb,Zhu03prb,Braig03prb,Flensberg03prb,Aji03condmat,Mitra04prb,Frederiksen04master,
Frederiksen04prl,GRN,Cizek05prb,Koch,Galperin06prb}. In parallel, the theory of inelastic
resonant tunneling in scanning tunneling spectroscopy was developed
\cite{Persson87prl,Gata93prb,MTU}.

The essential progress in calculation of transport properties in the strong
electron-vibron interaction limit has been made with the help of the master equation
approach \cite{Braig03prb,Mitra04prb,Koch}. This method, however, is valid only in the
limit of very weak molecule-to-lead couplings and neglects all spectral effects, which
are the most important at finite coupling to the leads.

Our approach is based on the nonequilibrium Green function
technique\,\cite{Kadanoff62book,Keldysh64,Rammer86rmp}, which is now a standard method in
mesoscopic physics and molecular electronics. We follow the formulation pioneered by
Meir, Wingreen, and Jauho\,\cite{Meir92prl,Jauho94prb,Haug96book}. This method was also
further developed in
Refs.\,\cite{Flensberg03prb,MTU,GRN,Frederiksen04master,Frederiksen04prl,Galperin06prb}.
The case of intermediate and strong electron-vibron couplings {\em at finite tunneling
rates} is the most interesting, but also the most difficult. Only the approaches by
Flensberg \cite{Flensberg03prb}, and Galperin, Ratner, and Nitzan \cite{Galperin06prb}
exist, both starting from the exact solution for the isolated system and then switching
on tunneling as a perturbation. Our approach is exact for non-interacting systems with
arbitrary tunneling coupling, and is approximate in the strong electron-vibron coupling
limit without tunneling. Still, its advantage is clear in the truly {\em nonequilibrium}
formulation.

The current in the left ($i=L$) or right ($i=R$) contact to the molecule is described by
the well-known expression
\begin{equation}\label{J}\begin{array}{c}\displaystyle
 J_{i=L,R}=\frac{\mathrm{i}e}{\hbar}\int\frac{d\epsilon}{2\pi}{\rm Tr}\left\{
 {\bf\Gamma}_i(\epsilon-e\varphi_i)\left({\bf G}^<(\epsilon)+ \right.\right.\\[0.5cm]
 \displaystyle \left.\left.
 f^0_i(\epsilon-e\varphi_i)
 \left[{\bf G}^R(\epsilon)-{\bf G}^A(\epsilon)\right]\right)\right\},
 \end{array}
\end{equation}
where $f^0_i(\epsilon)$ is the equilibrium Fermi distribution function in the lead
$i$ with the chemical potential $\mu_i$, and the level-width function
${\bf\Gamma}_{i=L(R)}(\epsilon)\equiv\Gamma_{i\alpha\beta}(\epsilon)$ is
\begin{equation}
\Gamma_{i\alpha\beta}(\epsilon) =2\pi\sum_{k\sigma}
V_{ik\sigma,\beta}V^*_{ik\sigma,\alpha}\delta(\epsilon-\epsilon_{ik\sigma}).
\end{equation}

The lesser (retarded, advanced) Green function matrix of a nonequilibrium molecule ${\bf
G}^{<(R,A)}\equiv G_{\alpha\beta}^{<(R,A)}$ can be found from the Dyson-Keldysh equations
in the integral form
\begin{eqnarray}
& {\bf G}^R(\epsilon)={\bf G}^R_0(\epsilon)
  +{\bf G}^R_0(\epsilon){\bf\Sigma}^R(\epsilon)
  {\bf G}^R(\epsilon), \\[0.3cm]
& {\bf G}^<(\epsilon)=
  {\bf G}^R(\epsilon){\bf\Sigma}^<(\epsilon){\bf G}^A(\epsilon),
\end{eqnarray}
or from the corresponding equations in the differential form (see
Refs.\,\onlinecite{Ryndyk05prb,Ryndyk06prb} and references therein).

Here ${\bf\Sigma}^{R,<}= {\bf\Sigma}^{R,<(T)}_{L}+{\bf\Sigma}^{R,<(T)}_{R}+
{\bf\Sigma}^{R,<(V)}$ is the total self-energy of the molecule composed of the tunneling
(coupling to the left and right leads) self-energies
${\bf\Sigma}_{j=L,R}^{R,<(T)}\equiv
 {\Sigma}_{j\alpha\beta}^{R,<(T)}=\sum_{k\sigma}\left\{V^*_{jk\sigma,\alpha}
 {G}^{R,<}_{jk\sigma}V_{jk\sigma,\beta}\right\}$,
and the vibronic self-energy ${{\bf\Sigma}}^{R,<(V)}\equiv
{\Sigma}^{R,<(V)}_{\alpha\beta}$. Here we assume, that vibrons are in equilibrium and are
not excited by the current, so that the self-consistent Born approximation is a good
starting point. The corresponding self-energies can be found in
Refs.\,\cite{Rammer86rmp,Haug96book,MTU,Mitra04prb,GRN,Frederiksen04master,Frederiksen04prl,Ryndyk05prb,Ryndyk06prb}.

\begin{figure}[t]
\begin{center}
\epsfxsize=1.0\hsize
\epsfbox{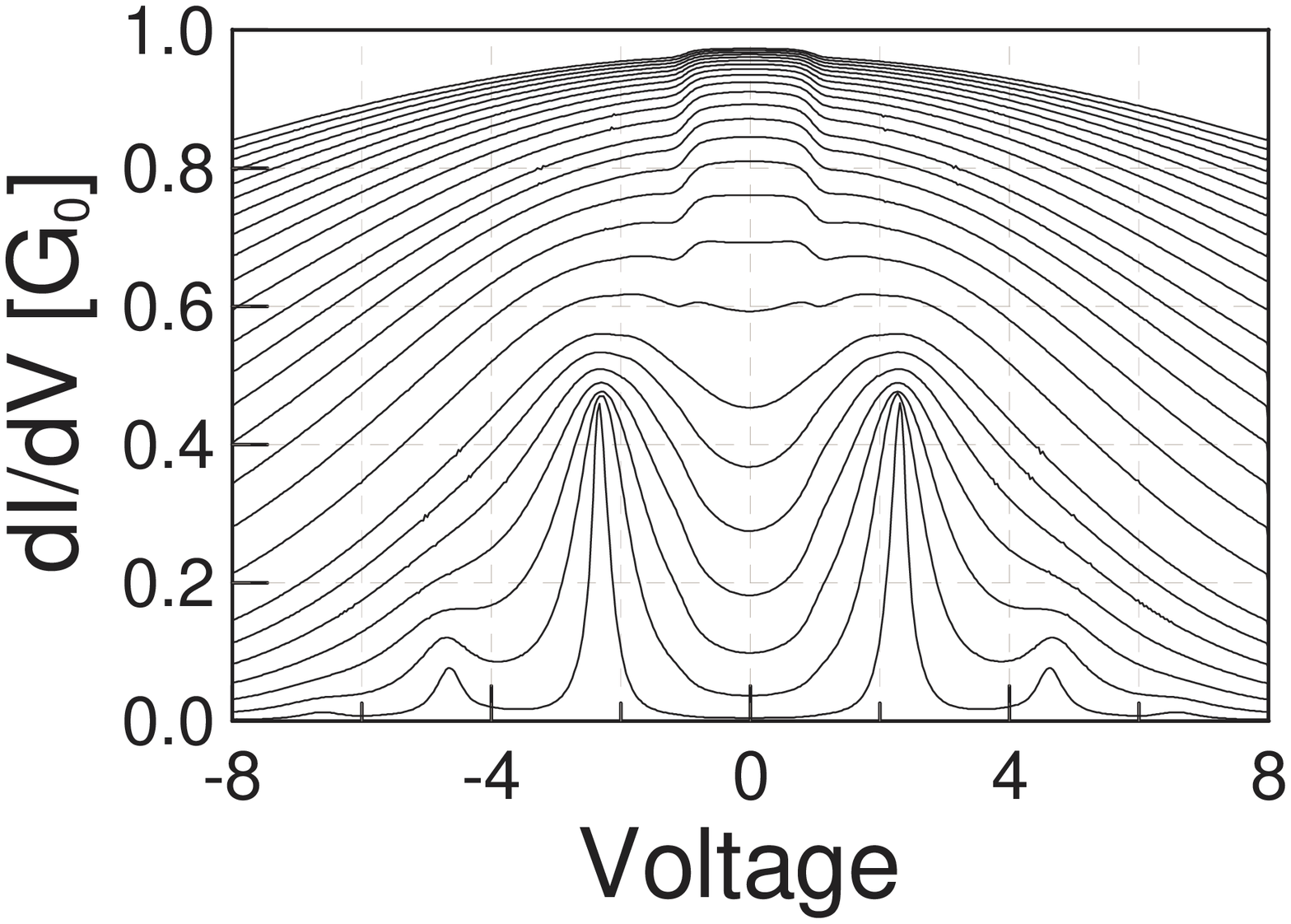}
\caption{Differential conductance of a {\em symmetric} junction ($\eta=0.5$, $\Gamma_R=\Gamma_L$) at different
molecule-to-lead coupling, from
\mbox{$\Gamma_L/\omega_0=0.1$} (lower curve) to $\Gamma_L/\omega_0=10$ (upper curve),
\mbox{$\lambda/\omega_0=1$}, $\epsilon_0/\omega_0=2$. Voltage is in the units of
$\hbar\omega_0$.}
 \label{fig3}
\end{center}
\end{figure}

For the single-level model all equations are significantly simplified. Combining $J_L$
and $J_R$ the expression for the current can be written for energy independent
$\Gamma_{L(R)}$ (wide-band limit) as
\begin{equation}\label{J_qc}
J=\frac{e}{2\pi\hbar}\frac{\Gamma_L\Gamma_R}{\Gamma_R+\Gamma_L} \int d\epsilon
A(\epsilon)\left[f^0(\epsilon-e\varphi_L)- f^0(\epsilon-e\varphi_R)\right].
\end{equation}
It looks as simple as the Landauer-B\"{u}ttiker formula, but it is not trivial,
because the spectral density $A(\epsilon)=-2{\rm Im}G^R(\epsilon)$ now depends on
the distribution function of electrons inside the fluctuating molecule and hence the
applied voltage, $\varphi_L=-\varphi_R=V/2$ \cite{Ryndyk05prb}. A typical example of
the spectral function at zero voltage is shown in Fig.~\ref{fig2}. At finite voltage
it is calculated self-consistently.

\begin{figure}[t]
\begin{center}
\epsfxsize=1.0\hsize
\epsfbox{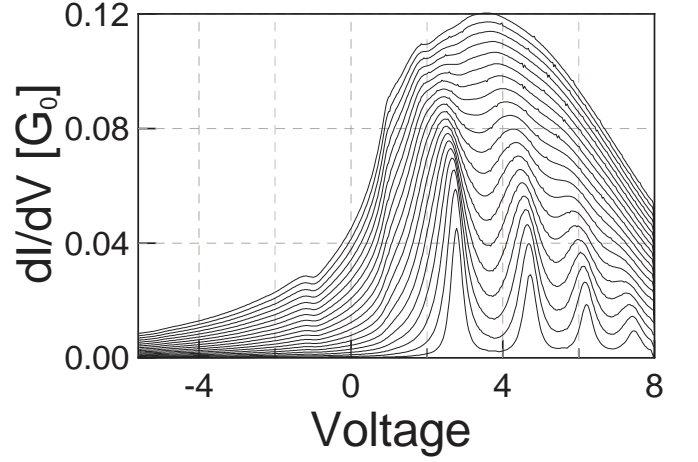}
\caption{Differential conductance of an {\em
asymmetric} junction ($\eta=0$, $\Gamma_R=20\Gamma_L$) at different molecule-to-lead
coupling, from \mbox{$\Gamma_R/\omega_0=0.2$} (lower curve) to $\Gamma_R/\omega_0=4$
(upper curve), $\lambda/\omega_0=2$, $\epsilon_0/\omega_0=5$. The voltage is in the units of
$\hbar\omega_0$}
\label{fig4}
\end{center}
\end{figure}

Before considering the experiments, let us discuss a general picture of the vibronic
transport in symmetric and asymmetric molecular bridges coupled to the electrodes,
provided in experiments by the MCBJ and STM technique, respectively. The differential
conductance, calculated at different molecule-to-lead coupling, is shown in
Fig.~\ref{fig3} (symmetric) and Fig.~\ref{fig4} (asymmetric). At weak coupling, the
vibronic side-band peaks are observed, reproducing the corresponding peaks in the
spectral function (Fig.~\ref{fig2}). At strong couplings the broadening of the electronic
state hides the side-bands, and new features become visible. In the symmetric junction, a
suppression of the conductance at $V\simeq\pm\hbar\omega_0$ takes place as a result of
inelastic scattering of the coherently transformed from the left lead to the right lead
electrons. In the asymmetric junction, the usual IETS increasing of the conductance is
observed at a negative voltage $V\simeq-\hbar\omega_0$, this feature is weak and can be
observed only in the incoherent tail of the resonant conductance. We conclude, that the
vibronic contribution to the conductance can be distinguished clearly in both coherent
and tunneling limits.

\begin{figure}[t]
\begin{center}
\epsfxsize=1.0\hsize
\epsfbox{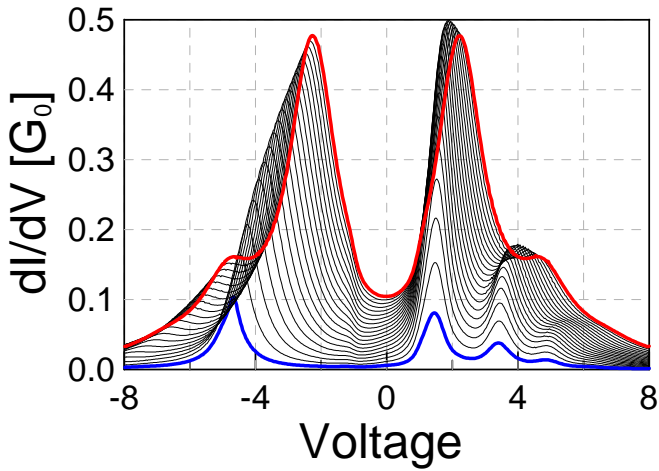}
\caption{(Color online) Differential conductance at different molecule-to-STM coupling
(see the text), from {\em asymmetric} junction with $\Gamma_L/\omega_0=0.025$,
$\Gamma_R/\omega_0=0.5$ and $\eta=0.2$ (lower curve, blue thick line) to
{\em symmetric} junction with $\Gamma_L/\omega_0=\Gamma_R/\omega_0=0.5$ and $\eta=0.5$
(upper curve, red thick line), \mbox{$\lambda/\omega_0=1$}, $\epsilon_0/\omega_0=2$.
Voltage is in the units of $\hbar\omega_0$}
\label{fig5}
\end{center}
\end{figure}

Now let us discuss the STS experiments \cite{Qiu04prl,Wu04prl,Repp05prl,Repp05prl2}. Here
we concentrate mainly on the dependence on the tip-to-molecule distance \cite{Wu04prl}.
When the tip (left lead in our notations) is far from the molecule, the junction is
strongly asymmetric: $\Gamma_L\ll\Gamma_R$ and $\eta\rightarrow 0$, and the conductance
is similar to that shown in Fig.\,\ref{fig4}. When the tip is close to the molecule, the
junction is approximately symmetric: $\Gamma_L\approx\Gamma_R$ and $\eta\approx 0.5$, and
the conductance curve is of the type shown in Fig.\,\ref{fig3}. We calculated the
transformation of the conductance from the asymmetric to symmetric case
(Fig.~\ref{fig5}). It is one new feature appeared in asymmetric case due to the fact that
we started from a finite parameter $\eta=0.2$ (in the Fig.~\ref{fig4} $\eta=0$), namely a
single peak at negative voltages, which is shifted to smaller voltage in the symmetric
junction. The form and behavior of this peak is in agreement with experimental results
\cite{Wu04prl}.

In conclusion, at weak molecule-to-lead (tip, substrate) coupling the usual vibronic
side-band peaks in the differential conductance are observed; at stronger coupling to the
leads (broadening) these peaks are transformed into step-like features. A
vibronic-induced decreasing of the conductance with voltage is observed in
high-conductance junctions. The usual IETS feature (increasing of the conductance) can be
observed only in the case of low off-resonant conductance. By changing independently the
bias voltage and the tip position, it is possible to determine the energy of molecular
orbitals and the spectrum of molecular vibrations. In the multi-level systems with strong
electron-electron interaction further effects, such as Coulomb blockade and Kondo effect,
could dominate over the physics which we address here; these effects have to be included
in a subsequent step.

We acknowledge fruitful discussions with Peter H\"anggi. This work was supported by the
Volkswagen Foundation under grant I/78~340, by the EU under contract IST-2001-38951, and
by the DFG CU 44/3-2.

\end{document}